\documentclass{iaus}
\usepackage{graphicx}

\title[PP.~~Gravitationally lensed quasars (GLQs)] 
{Time-domain studies of gravitationally lensed quasars (GLQs)}

\author[Luis J. Goicoechea \& Vyacheslav N. Shalyapin] 
{Luis J. Goicoechea $^1$ \and Vyacheslav N. Shalyapin$^{1,2}$
\thanks{On behalf of the GLENDAMA Project Team.}}

\affiliation{$^1$Dept. de F\'{\i}sica Moderna, Universidad de Cantabria, \\ Avda. de 
Los Castros s/n, ES-39005, Santander, Spain 
\\ email: {\tt goicol@unican.es} \\[\affilskip]
$^2$Inst. Radiophysics \& Elect., Nat. Acad. of Sci. of Ukraine, \\ 12 Proskura 
St., UA-61085, Kharkov, Ukraine
\\email: {\tt vshal@ukr.net}}

\pubyear{2012}
\volume{285}  
\pagerange{1--3}
\jname{New Horizons in Time Domain Astronomy}
\editors{R.E.M. Griffin, R.J. Hanisch \& R. Seaman, eds.}
\begin{document}

\maketitle

\begin{abstract}
We present the overview and current results of an ongoing optical/NIR monitoring of seven 
GLQs with the 2-m Liverpool Robotic Telescope. The photometric data over the first seven 
years of this programme (2005--2011) are leading to high-quality light curves, which in 
turn are being used as key tools for different standard and novel studies. While 
brightness records of non-lensed distant quasars may contain unrecognized extrinsic 
variations, one can disentangle intrinsic from extrinsic signal in certain GLQs. Thus, 
some GLQs in our sample allow us to assess their extrinsic and intrinsic variations, as 
well as to discuss the origin of both kinds of fluctuations. We also demonstrate the 
usefulness of GLQ time-domain data to obtain successful reverberation maps of inner 
regions of accretion disks around distant supermassive black holes, and to estimate 
redshifts of distant lensing galaxies. 
\keywords{gravitational lensing, black hole physics, accretion, galaxies: general, 
quasars: general.}
\end{abstract}


An overview of our ongoing Liverpool Quasar Lens Monitoring (LQLM) project is presented in 
Table \ref{tab1}. The data collection is being carried out in different phases: LQLM I 
(from January 2005 to July 2007), LQLM II (from February 2008 to July 2010) and LQLM III 
(from October 2010 to present), and using the available optical/NIR instrumentation. The 
relevant instruments on the Liverpool Robotic Telescope are the RATCam CCD camera and its 
associated Sloan $griz$ filter set, the RINGO2 optical polarimeter and the FRODOspec 
spectrograph (3900--9400 \AA). Some astrophysical results and expectations for each target 
(GLQ) follow below, while a more complete and updated information can be found on the 
GLENDAMA website {\tt http://grupos.unican.es/glendama}.

\begin{table}
  \begin{center}
  \caption{Current status of the LQLM project.}
  \label{tab1}
 {\scriptsize
  \begin{tabular}{|l|c|c|c|c|c|}\hline 
{\bf GLQ} & {\bf Comments$^1$} & {\bf Main lens} & {\bf Observation} & {\bf 
Instruments$^2$} & {\bf Outputs$^3$} \\ {\bf (redshift)} & & {\bf (redshift)} & {\bf 
phases$^2$} & & ({\bf status}) \\ \hline
SBS 0909+532 & 2 images: A-B & early-type galaxy & I+III & RATCam & LC \\ 
($z$ = 1.38) & size $\sim$ 1.11$^{\prime\prime}$ & ($z$ = 0.83) & & $gr$ filters 
& (final reduction) \\ \hline
FBQ 0951+2635 & 2 images: A-B & early-type galaxy & I+II+III & RATCam & LC+DI \\ 
($z$ = 1.25) & size $\sim$ 1.10$^{\prime\prime}$ & ($z$ = 0.26) & & $ri$ filters 
& (final reduction) \\ \hline
QSO 0957+561 & 2 images: A-B & cD galaxy & I+II+III & RATCam & LC+S+PM \\ 
($z$ = 1.41) & size $\sim$ 6.17$^{\prime\prime}$ & ($z$ = 0.36) & & $griz$ filters 
& (LC=final reduction, \\ & & & & FRODOspec & S=first reduction, \\ & & & & RINGO2 &
PM=pending) \\ \hline
SDSS 1001+5027 & 2 images: A-B & early-type galaxy & II & RATCam & LC \\ 
($z$ = 1.84) & size $\sim$ 2.86$^{\prime\prime}$ & ($z \sim$ 0.2--0.5) & & $g$ filter 
& (final reduction) \\ \hline
SDSS 1339+1310 & 2 images: A-B & early-type galaxy & II+III & RATCam & LC+DI \\ 
($z$ = 2.24) & size $\sim$ 1.69$^{\prime\prime}$ & ($z \sim$ 0.4) & & $ri$ filter 
& (first reduction) \\ \hline
HE 1413+117 & 4 images: A-D & ? & II & RATCam & LC+DI \\ 
($z$ = 2.56) & size $\sim$ 1.35$^{\prime\prime}$ & ($z$ = 1.9) & & $r$ filter 
& (final reduction) \\ \hline
QSO 2237+0305 & 4 images: A-D & face-on Sb galaxy & II & RATCam & LC \\ 
($z$ = 1.69) & size $\sim$ 1.78$^{\prime\prime}$ & ($z$ = 0.04) & & $gr$ filters 
& (first reduction) \\ \hline
  \end{tabular}
  }
 \end{center}
\vspace{1mm}
 \scriptsize{
 {\it Notes:}\\
  $^1$See the CASTLES ({\tt http://www.cfa.harvard.edu/castles/}) and \\ SQLS ({\tt 
  http://www-utap.phys.s.u-tokyo.ac.jp/~sdss/sqls/}) websites. \\
  $^2$See main text. \\
  $^3$LC = light curves, DI = deep images, S = spectra, PM = polarization measurements.}
\end{table}

{\underline{\it SBS 0909+532}}. The LQLM I light curves in the $r$ band led to a robust 
time delay between its two images of $\Delta t_{AB}$ = $-$49 $\pm$ 6 days ($\Delta t_{ij} 
= t_j - t_i$, B leading; \cite[Goicoechea et al. 2008a]{Goi08a}). In addition, the optical 
flux ratio $A/B$ changed little in the first 10 years of observations, i.e., between the 
identification as a quasar pair in 1996 and our LQLM I campaign (e.g., \cite[Dai \& 
Kochanek 2009]{Dai09} and references therein). For example, the $r$-band light curve of 
the A image and the properly shifted $r$-band light curve of B were consistent with each 
other throughout the LQLM I period, so the variability over this time segment was 
basically intrinsic to the distant quasar (\cite[Goicoechea et al. 2008a]{Goi08a}). 
However, the LQLM III light curves indicate that the $r$-band flux ratio has evolved in 
2010--2011. Gravitational microlensing by stars within the main lensing galaxy could 
account for the detected extrinsic variation.

{\underline{\it FBQ 0951+2635}}. Gravitational microlensing seems to be an important 
variability mechanism for this GLQ (e.g., \cite[Paraficz et al. 2006]{Par06}; 
\cite[Shalyapin et al. 2009]{Sha09}). We are taking a few frames per year in the $r$ 
band to trace the long-term behaviour of $A/B$, and thus, obtain information on the 
structure of the source and the lensing galaxy (e.g., \cite[Wambsganss 1990]{Wam90}; 
\cite[Kochanek 2004]{Koc04}).  

{\underline{\it QSO 0957+561}}. We did not find evidence of extrinsic variability in the 
LQLM I light curves in the $g$ and $r$ bands. These initial brightness records were used 
to measure time delays between images and optical bands (\cite[Shalyapin et al. 
2008]{Sha08}), as well as to analyse the structure function of the rest-frame UV 
variability (\cite[Goicoechea et al. 2008b]{Goi08b}; \cite[Goicoechea et al. 
2010]{Goi10}). Later, LQLM II fluxes in the $griz$ bands, together with concurrent 
space-based observations from Swift/UVOT and Chandra, unveiled details of the accretion 
flow and its jet connection in a distant radio-loud quasar for the first time (see 
contribution by Shalyapin, Goicoechea \& Gil-Merino, this volume, and \cite[Gil-Merino et 
al. 2011]{Gil11}). Our global database in the $gr$ bands is also providing surprising 
results on the chromaticity in $\Delta t_{AB}$ and the long-term evolution of $B/A$, which 
are likely related with the presence of a dense cloud within the cD lensing galaxy, along 
the line of sight to the A image. We are also exploring the spectro-polarimetric evolution 
of this fascinating, first GLQ. 

\begin{figure}[t]
\begin{center}
 \includegraphics[width=3.2in]{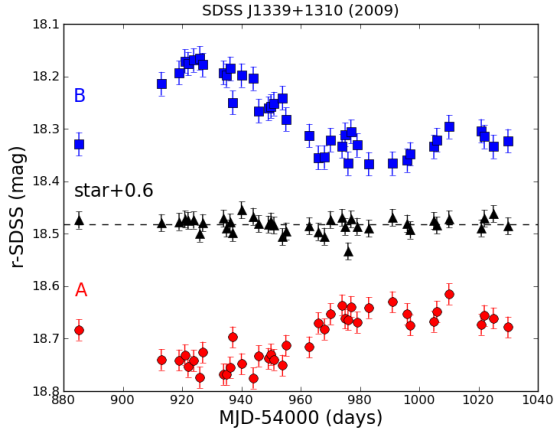} 
 \caption{LQLM II $r$-band light curves of SDSS 1339+1310.}
   \label{fig1}
\end{center}
\end{figure}

{\underline{\it Two new GLQs}}. SDSS 1001+5027 was discovered in 2005 (\cite[Oguri et al. 
2005]{Ogu05}), and the first monitoring campaign in the $R$ band did not produce any time 
delay between its two images (\cite[Paraficz et al. 2009]{Par09}). We have recently 
observed this double GLQ in the $g$ band (February-May 2010), since one expects to see 
more variability at shorter wavelengths. If A leads B, and there are no significant 
extrinsic variations, the LQLM II $g$-band fluxes suggest a time delay ranging from 12 to 
22 days. The other new GLQ (SDSS 1339+1310; \cite[Inada et al. 2009]{Ina09}) was monitored 
in the $r$ band just after its discovery (February-July 2009; LQLM II). Although we find 
prominent flux variations (Fig.\,\ref{fig1}), our LQLM II $r$-band light curves do not 
give any conclusive delay. Additional data in the LQLM III period are required to decide 
on the time delay and other properties of this system.

{\underline{\it Two famous quads}}. We followed up the $r$-band variability of the four 
images A-D of the Cloverleaf quasar (HE 1413+117) to measure its time delays for the first 
time. The LQLM II fluxes of this GLQ (February-July 2008) allowed us to obtain $\Delta 
t_{AB}$ = $-$17 $\pm$ 3, $\Delta t_{AC}$ = $-$20 $\pm$ 4 and $\Delta t_{AD}$ = 23 $\pm$ 4 
days (B-C leading, D trailing), which were used to estimate the redshift of the main 
lensing galaxies: $z$ = 1.88$^{+0.09}_{-0.11}$ (\cite[Goicoechea \& Shalyapin 
2010]{Gois10}). Although useful spectroscopic data are not yet available, we derived an 
accurate value of $z$ via gravitational lensing. We also monitored the Einstein Cross (QSO 
2237+0305) in the $g$ and $r$ bands, and the corresponding light curves and microlensing 
analyses will be presented soon. QSO 2237+0305 is the most emblematic target for 
microlensing studies (e.g., \cite[Shalyapin et al. 2002]{Sha02}; \cite[Kochanek 
2004]{Koc04}; \cite[Gil-Merino et al. 2006]{Gil06}).


\begin{thebibliography}{}

\bibitem[Dai \& Kochanek (2009)]{Dai09}
{Dai, X., \& Kochanek, C.S.} 2009, 
\textit{ApJ}, 692, 677

\bibitem[Gil-Merino et al. (2006)]{Gil06}
{Gil-Merino, R., Gonz\'alez-Cadelo, J., Goicoechea, L.J., Shalyapin, V.N., \& Lewis, G.F.} 
2006,
\textit{MNRAS}, 371, 1478

\bibitem[Gil-Merino et al. (2011)]{Gil11}
{Gil-Merino, R., Goicoechea, L.J., Shalyapin, V.N., \& Braga, V.F.} 2011,
\textit{ApJ}, in press (arXiv:1109.3330)
 
\bibitem[Goicoechea et al. (2008a)]{Goi08a}
{Goicoechea, L.J., Shalyapin, V.N., Koptelova, E., Gil-Merino, R., Zheleznyak, A.P., \& 
Ull\'an, A.} 2008a,
\textit{New Astron.}, 13, 182

\bibitem[Goicoechea et al. (2008b)]{Goi08b}
{Goicoechea, L.J., Shalyapin, V.N., Gil-Merino, R., \& Ull\'an, A.} 2008b,
\textit{A\&A}, 492, 411

\bibitem[Goicoechea et al. (2010)]{Goi10}
{Goicoechea, L.J., Shalyapin, V.N., \& Gil-Merino, R.} 2010,
\textit{The Open Astron. J.}, 3, 193 (see {\tt http://www.benthamscience.com/open/toaaj/})

\bibitem[Goicoechea \& Shalyapin (2010)]{Gois10} 
{Goicoechea, L.J., \& Shalyapin, V.N.} 2010,
\textit{ApJ}, 708, 995
 
\bibitem[Inada et al. (2009)]{Ina09}
{Inada, N., et al.} 2009,
\textit{AJ}, 137, 4118

\bibitem[Kochanek (2004)]{Koc04}
{Kochanek, C.S.} 2004, 
\textit{ApJ}, 605, 58

\bibitem[Oguri et al. (2005)]{Ogu05}
{Oguri, M., et al.} 2005,
\textit{ApJ}, 622, 106

\bibitem[Paraficz et al. (2006)]{Par06}
{Paraficz, D., Hjorth, J., Burud, I., Jakobsson, P., \& El\'{\i}asd\'ottir, \'A.} 2006,
\textit{A\&A}, 455, L1

\bibitem[Paraficz et al. (2009)]{Par09}
{Paraficz, D., Hjorth, J., \& El\'{\i}asd\'ottir, \'A.} 2009,
\textit{A\&A}, 499, 395

\bibitem[Shalyapin et al. (2002)]{Sha02}
{Shalyapin, V.N., Goicoechea, L.J., Alcalde, D., Mediavilla, E., Mu\~noz, J.A., \& 
Gil-Merino, R.} 2002,
\textit{ApJ}, 579, 127

\bibitem[Shalyapin et al. (2008)]{Sha08}
{Shalyapin, V.N., Goicoechea, L.J., Koptelova, E., Ull\'an, A., \& Gil-Merino, R.} 2008,
\textit{A\&A}, 492, 401

\bibitem[Shalyapin et al. (2009)]{Sha09}
{Shalyapin, V.N., et al.} 2009,
\textit{MNRAS}, 397, 1982

\bibitem[Wambsganss (1990)]{Wam90}
{Wambsganss, J.} 1990,  
 \textit{PhD thesis}, Munich University (also available as report MPA 550)

\end{thebibliography}
\end{document}